\documentclass{iopart}
\usepackage{iopams}

 \newcommand{\beq}{\begin{eqnarray}}
\newcommand{\eeq}{\end{eqnarray}} \newcommand{\be}{\begin{eqnarray}}
\newcommand{\ee}{\end{eqnarray}} 

\newcommand{\Oo}{{\cal O}}
\newcommand{\dd}{{\rm d}}

\begin{document}

\title{Indistinguishable Macroscopic Behaviour of Palatini  Gravities and General Relativity}

\author{Baojiu Li$^1$, David F. Mota$^2$ and Douglas J. Shaw$^3$}
\address{$^1$DAMTP, Centre for Mathematical Sciences, University
of Cambridge, Wilberforce Road, Cambridge CB3 0WA, UK}
\address{$^2$Institut f\"ur Theoretische Physik, Universit\"at
Heidelberg,  D-69120 Heidelberg, Germany}
\address{$^3$School of Mathematical Sciences, Queen Mary, University of London, London E1 4NS, UK}

\begin{abstract}
We show that, within some modified gravity theories, such as the
Palatini models, the non-linear nature of the field equations
implies that the usual na\"{i}ve averaging procedure (replacing
the microscopic energy-momentum  by its cosmological average)
could be invalid. As a consequence, the relative motion of
particles in Palatini theories is actually indistinguishable from
that predicted by General Relativity. Moreover, there is no WEP
violation. Our new and most important result is that the cosmology
and astrophysics, or put more generally, the behaviours on
macroscopic scales, predicted by these two theories are the same,
and as a result the naturalness problems associated with the
cosmological constant are not alleviated. Palatini gravity does
however predict alterations to the internal structure of particles
and the particle physics laws, \emph{e.g.}, corrections to the
hydrogen energy levels. Measurements of which place strong
constraints on the properties of viable Palatini gravities.
\end{abstract}

\maketitle

Extensions of General Relativity (GR) have always received a great
deal of attention. Such theories are motivated by quantum gravity
models and by the wish to find phenomenological alternatives to
the standard paradigm of dark matter and dark energy
\cite{carroll}.

Modifications to Einstein gravity generally result in non-linear
(in energy momentum tensor $T_{\mu\nu}$) corrections to the field
equations. The application of these equations equations to
macroscopic (\emph{e.g.} cosmological) scales involves an implicit
coarse-graining over the microscopic structure of particles.
However, when there are extra non-linear terms in the field
equations, \emph{a priori}, the validity of the usual
coarse-graining procedure can no longer be taken for granted, as
first proposed by \cite{flan}. Hence, as discussed in \cite{flan},
it may be important to take into account the microscopic structure
of matter when applying the field equations to macroscopic scales.

Unfortunately, to date, that has not been the common practice
\cite{pal1}. This is probably because, in GR, as in Newtonian
gravity, the microscopic structure of matter is not particularly
important on macroscopic scales. It is standard practice to
replace the metric, $g_{\mu\nu}$, and the energy momentum tensor,
$T_{\mu\nu}$, with some average of them that coarse-grains over
the microscopic structure. This simple procedure only works,
however, because on microscopic scales the equations of GR are
approximately linear.

In this \emph{Letter} we show that such an approach \emph{cannot}
simply be applied to modified gravity without a detailed first
analysis of the energy-momentum microstructure. Indeed,
na\"{i}vely averaging over the microscopic structure will
generally lead one to make incorrect predictions, and inaccurate
conclusions as to the validity of the theory \cite{flan}. Indeed,
it is possible that a theory deviates \emph{significantly} from GR
at the level of the microscopic field equations to be
\emph{indistinguishable} from the latter when correctly coarse
grained over macroscopic, \emph{e.g.}, cosmological, scales.

We illustrate this point for a class of modified gravity theories in
which the Ricci scalar, $R$, in the Einstein-Hilbert action is
replaced by some function $f(R, R^{\mu\nu}R_{\mu\nu})$,
\emph{i.e.}:
$$
\frac{1}{2\kappa}\int {\rm d}^4\,x \,\sqrt{-\bar{g}}R \rightarrow
\frac{1}{2\kappa}\int {\rm d}^4\,x \,\sqrt{-\bar{g}}f(R,
R^{\mu\nu}R_{\mu\nu}).
$$
The field equations for this action can be derived according to
two \emph{inequivalent} variational approaches: metric and
Palatini. In the former, $R_{\mu\nu}$ and $R$ are taken to be
constructed from the matter metric $\bar{g}_{\mu\nu}$, which
couples to matter and governs the conservation of energy momentum
tensor, and the field equations are found by minimizing the action
with respect to variations in $\bar{g}_{\mu \nu}$. In the
alternative, Palatini approach, $R = R_{\mu\nu}\bar{g}^{\mu\nu}$
where $R_{\mu\nu}$ is a function of some connection field
$\Gamma^{\mu}_{\nu\rho}$ which is, \emph{a priori}, treated as
being independent of $\bar{g}_{\mu\nu}$. The field equations are
then found by minimizing the action with respect to both
$\Gamma^{\mu}_{\nu\rho}$ and $\bar{g}_{\mu\nu}$. If $f(R,
R^{\mu\nu}R_{\mu\nu}) = R-2\Lambda$ (\emph{i.e.} GR with a
cosmological constant) then the two approaches result in the same
field equations. Otherwise they are generally different. Note that
the Palatini $f(R)$ field equations are mathematically equivalent
to an $\omega = -\frac{3}{2}$ Brans-Dicke theory with a potential,
and thus represents more general modified gravity theories.


Within the metric approach to the $f(R)$ gravity theories,
averaging over \emph{microscopic} scales is generally no less
straightforward than it is in GR; this is because in both cases
all degrees of freedom are dynamical. These dynamics normally
ensure that the field equations, for all degrees of freedom, are
approximately linear (in energy momentum tensor $T_{\mu\nu}$) on
small-scale structures \cite{longversion}. In contrast, averaging
in Palatini models is not so trivial as the new degree of freedom
is \emph{non-dynamical}, and so its field equation remains
non-linear (in energy momentum tensor $T_{\mu\nu}$) even on the
smallest scales. Consequently the cosmological behaviour of these
theories can be very different from what has been suggested in the
literature \cite{pal1, olmo, pal4}, which adopt the same averaging
procedure as in the metric approach. Also, although GR plus a
cosmological constant is a special case of Palatini $f(R)$ (and
also of metric $f(R)$!) theories, the averaging problem for
general Palatini theories dose not arise there because the GR
field equations are linear (and algebraic) in curvature $R$ and
energy momentum tensor components, while in contrast for general
Palatini theories the curvature depends nonlinearly (but also
algebraically) on energy momentum tensor. We emphasize again that
it is \emph{this} "nonlinearity" that makes the averaging of
general Palatini theories not as trivial as that in GR.

Palatini $f(R^{\mu\nu}R_{\mu\nu})$ theories are similar to the
$f(R)$ ones in many aspects, but their study is more complicated.
In what follows we shall mainly focus on the latter and state the
results of the former when appropriate, referring more details to
{\cite{longversion}}. To avoid confusion later, we now take $R
\rightarrow \mathcal{R}$, where $\mathcal{R} = R_{\mu
\nu}(\Gamma)\bar{g}^{\mu\nu}$.

Varying the $f(\mathcal{R})$ action with respect to
$\Gamma^{\mu}_{\nu\rho}$ gives that $\Gamma^{\mu}_{\nu\rho}$ is
the Levi-Civit$\mathrm{\grave{a}}$ connection of $g_{\mu\nu} =
f^{\prime}(\mathcal{R}) \bar{g}_{\mu\nu}$.  Varying the action
with respect to $\bar{g}_{\mu \nu}$ gives:
\begin{eqnarray}
G^{\mu}{}_{\nu}(g) &=& \frac{1}{f^{\prime\,2}(\Phi)}\left[\kappa
\mathcal{T}^{\mu}{}_{\nu}- V(\Phi)\delta^{\mu}{}_{\nu}\right], \label{eqn1}\\
\kappa {\mathcal T}^{\mu}{}_{\mu} &\equiv& \kappa \mathcal{T}\ =\
f^{\prime}(\Phi)\Phi - 2f(\Phi), \label{eqn2}
\end{eqnarray}
where we have defined $\Phi \equiv \mathcal{R}$, $R \equiv R_{\mu
\nu}g^{\mu \nu} = \Phi/f^{\prime}(\Phi)$, $G_{\mu \nu}(g) = R_{\mu
\nu} - \frac{1}{2}g_{\mu \nu} R$ and:
$$
V(\Phi) = \frac{1}{2}\left(f^{\prime}(\Phi)\Phi -
f(\Phi)\right),\qquad \mathcal{T}^{\mu}{}_{\nu} =
-\frac{2\bar{g}^{\mu \rho}}{\sqrt{-\bar{g}}} \frac{\delta S_{\rm
matter}}{\delta \bar{g}^{\rho \nu}}.
$$
Also $\nabla^{(\bar{g})}_{\mu}\mathcal{T}^{\mu}{}_{\nu} = 0$,
where $\nabla^{(\bar{g})}_{\mu} \bar{g}_{\nu \rho} = 0$ defines
$\nabla^{(\bar{g})}_{\mu}$. In $f(R^{\mu\nu}R_{\mu\nu})$ gravity
the field equation is similar to Eq.~(\ref{eqn1}), but the metrics
$g_{\mu\nu}$ and $\bar{g}_{\mu\nu}$ are \emph{disformal}, and $V$
depends \emph{algebraically} on $\mathcal{T}_{\mu\nu}$ rather than
simply on $\mathcal{T}$ \cite{longversion}.

It is important to bear in mind that Eqs.~(\ref{eqn1}) and
(\ref{eqn2}) are \emph{microscopic} field equations, and they are
only definitely valid when all the microscopic structure in the
distribution of energy and momentum is taken into account.
Nonetheless, the cosmological and astrophysical behaviours of
these theories have, to date, been studied by simply replacing
$g_{\mu\nu}$, $\Phi$ and $\mathcal{T}^{\mu}{}_{\nu}$ in
Eqs.~(\ref{eqn1}, \ref{eqn2}) by some coarse-grained averages of
them \cite{olmo, pal1, pal4}. However, in Palatini
$f(\mathcal{R})$ theories, a deviation from GR requires that $f$
depends \emph{nonlinearly} on $\Phi$. This nonlinearity introduces
an averaging problem for the Palatini theories. As we shall show,
this means that the standard averaging procedure is no longer
valid and generally results in incorrect physical and mathematical
predictions. Furthermore, when the microscopic structure of matter
is taken into account, the late-time cosmology of essentially
\emph{all} $f(\mathcal{R})$ Palatini theories is indistinguishable
from that of standard GR with a cosmological constant. We also
show that these these theories do \emph{not} violate the weak
equivalence principle (WEP). Palatini $f(R)$ gravity \emph{is}
however a different theory from GR and does produce alterations to
microscopic particle physics that could be detected.

The microscopic structure of the space-time distribution of matter
energy density, $\rho$, will not affect the macroscopic, or
coarse-grained, behaviour of the theory if and only if the field
equations of the theory are \emph{linear in $\rho$}. Here is a
simple example: consider a region of space with average density
$\left\langle \rho \right \rangle$ and volume $\mathcal{V}$, which
contains $N$ particles each with density $\rho_{c}$ and volume
$\mathcal{V}_{p}$. The space in between the particles is empty and
so $\left\langle \rho \right \rangle \mathcal{V} = N \rho_c
\mathcal{V}_{p}$. Now consider some quantity $Q(\rho)$.  Inside
the particles, $\rho = \rho_c$ and so $Q = Q_c \equiv Q(\rho_c)$;
outside $\rho = 0$ and so $Q = Q_0 \equiv Q(0)$.  The average
value of $Q$ (by volume) is:
\begin{equation}
\left\langle Q \right\rangle =
Q_{0}\left(1-\frac{N\mathcal{V}_p}{\mathcal{V}}\right) +
Q_{c}\frac{N \mathcal{V}_p}{\mathcal{V}} = Q_{0} + \frac{Q_c -
Q_0}{\rho_c} \left\langle \rho \right \rangle \label{avgeqn}
\end{equation}
It is clear that, irrespective of how $Q$ depends on $\rho$,
$\left\langle Q\right \rangle $ depends on $\left\langle \rho
\right \rangle$ linearly. Thus if $Q$ depends nonlinearly on
$\rho$, we have $\left \langle Q(\rho) \right\rangle \neq
Q\left(\left\langle \rho \right\rangle\right)$. In Palatini
theories, $V(\Phi)$ generally exhibits highly non-linear
dependence on $\rho$, and so $\left\langle
V(\Phi(\rho))\right\rangle \neq V(\Phi(\left\langle \rho
\right\rangle))$.

It is also interesting to know when na\"{i}vely coarse-grained
equations are valid to a good approximation over the length scales
and density scales where we have observations. This only occurs in
Palatini theories when the inherently nonlinear modifications to
GR have sub-leading effect. In all these cases the theory will
reduce to GR up to the order where the na\"{i}vely coarse-grained
equations are valid. In this work we are concerned with the how
Palatini theories deviate from GR, even when those deviations are
sub-leading order. We must then always take account of those
nonlinear terms in the equations which determine how Palatini
theories deviate from GR. The crucial r\^{o}le played by these
nonlinear terms means that the na\"{i}ve coarse-graining procedure
will always fail to accurately describe the differences between
Palatini theories and GR.

To uncover how Palatini gravities behave on macroscopic scales, we
consider how a set of microscopic particles evolve under such a
modified gravity. Consider a single, spherical particle for which
$\mathcal{T}^{\mu}_{\nu} \neq 0$ for $R < R_{p}$ but vanishes
otherwise.  The metric for such a particle is  \cite{olmo}:
$$
g_{\mu \nu}\dd x^{\mu} \dd x^{\nu} = -W(r) e^{2\chi(r)} \dd t^2  +
\frac{1}{W(r)} \dd r^2 + r^2 \dd \Omega^2,
$$
with
\begin{eqnarray}
W(r) &=& 1 - \frac{2GM(r)}{r}-\frac{V(\Phi_0)}{3 f^{\prime\,2}(\Phi_0)}r^2, \\
2GM(r) &=& \kappa\int_{0}^{r}\dd
r^{\prime}\,r^{\prime\,2}\left[\frac{\rho}{f^{\prime\,2}(\Phi)}
+ \frac{\Delta V(\Phi)}{\kappa f^{\prime\,2}(\Phi)}\right], \label{Gmass} \\
\chi &=&\frac{\kappa}{2}\int_{R_p}^{r} \dd r^{\prime}\,
r^{\prime}\frac{(\rho + p_I+p_{A})}{W(r^{\prime})
f^{\prime\,2}(\Phi)},
\end{eqnarray}
where $$\Delta V \equiv V(\Phi) -
f^{\prime\,2}(\Phi)V(\Phi_0)/f^{\prime\,2}(\Phi_0),$$ $$
\mathcal{T}^{\mu}_{\ \ \nu} = \left(\rho+p_{I}\right)
u^{\mu}u^{\rho}\bar{g}_{\rho \nu} + p_{I}\delta^{\mu}{}_{\nu} +
\pi^{\mu}{}_{\nu},$$ $u^{\mu}u^{\nu}\bar{g}_{\mu\nu} = -1$ and
$\pi^{\mu}{}_{\nu}$ is the anisotropic stress satisfying
$\pi^{\mu}{}_{\nu}u^{\nu} = 0$. In the rest frame of the particle
$$\pi^{\mu}{}_{\nu} = {\rm diag}(0, p_{A}, -p_{A}/2, -p_{A}/2), \qquad \Phi_0 \equiv \Phi(\mathcal{T} = 0).$$ The presentation of this
solution in Ref.~\cite{olmo} was, however, incomplete as it failed
to note that the $\theta \theta$ component of Eq.~(\ref{eqn1})
results in the condition:
\begin{equation}\label{TOV}
r\frac{\dd}{\dd r}P_{\rm eff} + \frac{\rho + p_I +
p_A}{f^{\prime\,2}(\Phi) rW(r)}Y(r) =
-3\frac{p_{A}}{f^{\prime\,2}(\Phi)} \label{volk}
\end{equation}
where
$$Y(r) = \left(4\pi G P_{\rm eff}r^3 + GM -
\frac{\Lambda_{\rm eff}r^3}{3} \right),$$ $\Lambda_{\rm eff} =
V(\Phi_0)/f^{\prime\,2}(\Phi_0)$ and $$P_{\rm eff} =
(p_{I}+p_A-\Delta V/\kappa)/f^{\prime\,2}.$$ Eq.~(\ref{volk})
implies that at $R=R_{p}$ one has $P_{\rm eff} = 0$.

Outside the particle, $\chi = 0$ and $$W(r) = 1-GM_p/r
-\Lambda_{\rm eff}r^2/3$$ where $M_{p} = M(R_p) = {\rm const}$ and
so the metric is precisely that of a Schwarzschild de-Sitter
spacetime with gravitational mass $M_{p}$ and an effective
cosmological constant $\Lambda_{\rm eff}$. A similar Schwarzschild
de-Sitter solution is found for the Palatini
$f(R^{\mu\nu}R_{\mu\nu})$ case \cite{longversion}, with $V$
replaced by a more complicated function of $\mathcal{T}^{\mu}_{\ \
\nu}$. Thus we see that in Palatini models the external metric
of a single particle is precisely what it would be in General Relativity with a
cosmological constant.

We now consider a spacetime containing many such particles. We
define $\mathbf{x}_{(I)}(t)$ to be the position of a particle $I$
and $v^2_{(I)} = \dot{x}_{(I)}^2$. Making a weak field
approximation with respect to $g_{\mu\nu}$ and using the above
solution, we find that inside the particle labeled $K$ and to
$\Oo(\epsilon^2)$, where $\epsilon \sim {\rm max}(\vert
v_{(K)}\vert, \sqrt(1-W(r)))$:
\begin{eqnarray}
g_{\mu \nu}\dd x^{\mu} \dd x^{\nu} &=&
-\left(W(r)+2\chi(r)-2U(x)\right)\dd t^2+
\left(1+2U(x)\right)\dd x^{k} \dd x^{k}  \nonumber \\  &+& \frac{(1-W(r))}
{r^2}\Delta x^{i}_{(K)} \Delta x^{j}_{(K)}\dd x^{i} \dd x^{j}, \nonumber \\
U(x) &=& \sum_{I\neq K} \frac{Gm_{A\,(I)}}{\vert
x^{i}-x_{(I)}^{i}(t)\vert}, \nonumber
\end{eqnarray}
$\Delta x_{(K)}^{i} = x^{i}-x^{i}_{(K)}(t)$ and $r = \vert
\Delta x_{(K)}^{i} \vert$; $m_{A\,(I)}$ is the active
gravitational mass of each particle given by $M(R_p)$.

Let $u^{\mu}_{(K)}$ be the 4-velocity of the $K^{\rm th}$ particle
which satisfies $u^{\mu}_{(K)}u^{\nu}_{(K)}\bar{g}_{\mu \nu}=-1$.
To the order $\Oo(\epsilon)$, $$u^{\mu}_{(K)} =
f^{\prime\,1/2}(\Phi)\left(1,\dd x^{i}/\dd t\right).$$ Now
$\mathcal{T}^{\mu}{}_{\nu}$ is conserved with respect to
$\bar{g}_{\mu \nu}$ \emph{i.e.}
$\nabla^{(\bar{g})}_{\mu}\mathcal{T}^{\mu}{}_{\nu}=0$. Evaluating
this equation at the centre of the particle we have $u_{K}^{\mu}
\nabla^{(\bar{g})}_{\mu}u_{K}^{\nu} = 0$, \emph{i.e.}
\begin{equation}
\label{eu}
\frac{\dd^2 x_{(K)}^{i}}{\dd \tau^2} =  \frac{1}{2}\bar{g}^{ij}
\bar{g}_{00,j} u^{0}u^{0} = - f^{\prime}(\Phi) U_{,i}.
\end{equation}
where $\tau$ is the proper time along the worldline of the
particle, and so $\partial t / \partial \tau = f^{\prime\,1/2}(\Phi)$
(because the metric $-d\tau^{2} =
\frac{1}{f'^{2}}(-dt^{2}+d\mathbf{x}^{2})$ means
$1=\frac{1}{f'^{2}}\left[(dt/d\tau)^{2}+(d\mathbf{x}/d\tau)^{2}\right]
\approx\frac{1}{f'^{2}}(dt/d\tau)^{2}$). 

The internal
configuration of the particle can be static or non-static, and
here for simplicity we assume it to be static and will comment on
the non-static case below. This requires that all gradients in
$\Phi$ cancel with gradients in the pressure and also that
$\mathcal{T}$ and hence $\Phi$ are conserved along particle
worldlines, \emph{i.e.}, $u^{\mu}\partial_{\mu} \mathcal{T} =
u^{\mu}\partial_{\mu} \Phi = 0$. Equation (\ref{eu}) is then
equivalent to:
\begin{equation}
a_{(K)}^{i} \equiv \frac{\dd^2 x_{(K)}^{i}}{\dd t^2} = - \left.
U_{,i}\right\vert_{\mathbf{x}=\mathbf{x}_{(K)}(t)}.
\end{equation}
The acceleration with respect to $\tau$ depends both on the
gravitational field, $U_{,i}$, and $f^{\prime}(\Phi)$. The
acceleration measured with respect to $t$, however, depends only
on $U_{,i}$.  The relative acceleration of two particles labeled
$1$ and $2$ as measured by a third, labeled $3$ say, with proper
time  $\tau_{(3)}$ is therefore:
$$
\Delta a^{i}_{12} = \frac{\dd^2 (x_{(1)}^{i}-x_{(2)}^{i})}{\dd^2
\tau_{(3)}} = f^{\prime}(\Phi_{(3)})
\left(a_{(1)}^{i}-a_{(2)}^{i}\right),
$$
where $\partial t/\partial \tau_{(3)} =
f^{\prime\,1/2}(\Phi_{(3)})$. In a uniform gravitational field:
$U_{,i} = {\rm const}$ and so $$a_{(1)}^{i}=a_{(2)}^{i}
\Rightarrow \Delta a^{i}_{12} = 0.$$  It follows that an observer
(\emph{i.e.}~particle $3$) sees any two other particles, $1$ and
$2$, accelerate at the same rate in a uniform gravitational field.
This is precisely what is required by the Weak Equivalence
Principle. 

Since the internal configuration of
the particles is static and the centres of two particles do not have
relative acceleration, there will be no relative acceleration
between the two whole particles. In Ref.~\cite{olmo} it was
suggested that internal gradients in $\Phi$ would lead to WEP
violations. In fact, because those gradients all vanish outside
the particle, they cannot affect the overall motion of the
particle \cite{flanmotion} (this is related to the fact
that the extra force, which arises due to the modification to GR, has zero
range; as we shall discuss in more detail below). In this case,
hydrostatic equilibrium, Eq.~(\ref{volk}), ensures that gradients
in $\Phi$ are cancelled by pressure gradients.

The absence of WEP violation in the dynamics of particles ensures
that the inertial and passive gravitational mass of particles are
equal. Moreover, the inertial mass and the active gravitational
mass of particles are also equal in these theories. Let us define
\begin{equation}
\label{eins} \kappa T_{{\rm eff}\,\nu}^{\mu} = G^{\mu}{}_{\nu}(g)
+ \Lambda_{\rm eff} \delta^{\mu}{}_{\nu}.
\end{equation}
By the modified Einstein equation, Eq.~(\ref{eqn1}), and the
contracted Bianchi identity: $\nabla^{(g)}_{\mu}T^{\mu}{}_{{\rm
eff}\, \nu} = 0$ where $\nabla^{(g)}_{\mu} g_{\nu \rho} = 0$.
Outside an isolated particle $T^{\mu}{}_{{\rm eff}\,\nu} = 0$ and,
as we have seen above, the metric $g_{\mu \nu}$ outside this
quasi-static (\emph{i.e.} $v^{2}_{I} \ll 1$) isolated particle is
Schwarzschild-de-Sitter. In Ref.~\cite{tolman}, Tolman shows that,
under these conditions, the inertial and active gravitational mass
of an isolated (not necessarily spherically symmetric) system are
equal (see also \cite{flanmotion}). For our particles, the
inertial, passive and active gravitational mass are therefore
equal to $m_{p}$ where by Eq.~(\ref{Gmass}):
\begin{equation}
m_{p} = 4\pi \int_{0}^{R_{p}} r^2 \dd r \left[\frac{\kappa\rho +
\Delta V}{\kappa f^{\prime\,2}(\Phi)}\right].
\end{equation}
This equivalence holds for particles both in GR and in Palatini
$f(\mathcal{R}, R_{\mu\nu}R^{\mu\nu})$ theories
\cite{longversion}. It follows that the motion of isolated
particles (situated in a vacuum) in Palatini theories is exactly
the same as it is in GR with $\Lambda = \Lambda_{\rm eff}$. This
has a very simple physical explanation. In all modified gravity
theories one may think of the modifications as being due to some
new, effective force. In Palatini theories, there is no extra
dynamical degree of freedom, so this `new' force is also
non-dynamical, \emph{i.e.}, does not propagate and only acts at
points. The closest analogue to this in particle physics would be
Fermi's original proposal for a theory of the weak interaction.
The effective new force is entirely \emph{local} and depends on
gradients in the $\mathcal{T}$ or $\mathcal{T}^{\mu}_{\ \nu}$.
There is therefore no way for one isolated particle to influence
the motion of another, and the force simply vanishes in a vacuum.
The effective force will alter the internal configuration of the
particle nonetheless, but this will not effect the motion of the
particle as a whole. Note that the above discussion relies on the
specific model for the matter distribution in a system. In most
systems it is realistic to take the matter as to be distributed in
localized clumps (what we refer to as particles) which only
interact with each other gravitationally (or through extra forces
due to the modification of gravity, which as we said above is of
zero range and thus completely irrelevant for classical
particles). For the extreme environments like a neutron star,
however, the energy density is of order nuclear density, and there
is essentially no space between different particles so that the
above matter model breaks down: in this case we expect the
na\"{\i}ve averaging procedure to be applicable. For cosmology and
most other astrophysical systems our averaging is nonetheless more
realistic. We emphasize that the absence of WEP violation does not require the 
internal configuration of the particles is static (as we have assumed). Rather, it is a
result of the fact that the extra force has zero range so that
there is no composition-dependent extra force between different
particles which do not overlap with each other. Additionally Palatini $f(R)$ theories are equivalent to a generalized Brans-Dicke, and, in common with other Brans-Dicke theories, the extra force has no composition dependence.

We now apply our results to a cosmological setting. At late times
and on a microscopic level, most matter in the Universe is made up
of small particles and a bit of radiation. In $f(\mathcal{R})$
theories, $\mathcal{T} = 0$ for radiation and so the presence of
radiation does not alter the relation between $\Phi$ and
$\mathcal{T}$. Our analysis is therefore directly applicable to
this setting. The Universe must therefore evolve precisely as it
would in GR with a cosmological constant $\Lambda_{\rm eff}$. This
can be seen in an alternative manner: averaging Eq.~(\ref{eins})
over volume $\mathcal{V}$ containing $N$ particles with mass
$m_{p}$ and using Eq.~(\ref{avgeqn}) we find
$$
\kappa\rho_{\rm matter}^{\rm eff} = -\left\langle \kappa T_{{\rm eff}\,0}^{0} \right \rangle = \kappa m_{p}\frac{N}{\mathcal{V}}.
$$
Using Eq.~(\ref{volk}) we find that in the rest frame of the
particle $\left\langle \kappa T^{i}_{{\rm eff}\,j} \right \rangle=
0$. More generally then $\left\langle \kappa T^{i}_{{\rm eff}\,j}
\right \rangle \sim \mathcal{O}(\kappa \rho_{\rm matter}^{\rm eff}
\delta v^{i} \delta v_{j})$ where $\delta v^{i}$ is the relative
particle velocity. Thus, when correctly coarse-grained over
cosmological scales, $T^{\mu}_{{\rm eff}\,\nu}$ describes a
collisionless dust; we have assumed that the peculiar velocities,
$\delta v$, of the particles are small and dropped terms of
$\mathcal{O}(\delta v^2)$. The cosmological evolution of the
Universe in such theories is therefore precisely the same as it is
in GR with a cosmological constant $\Lambda_{\rm eff}$ and a dust
with energy density $\rho_{\rm matter}^{\rm eff}$. This argument
also holds for most astrophysical systems. In particular, solar
system tests are evaded and the Parametrized Post-Newtonian
parameters are indistinguishable from those of GR.

Many of the predictions made in the literature (\emph{e.g.}
\cite{pal1}) do not, therefore, follow from a Palatini
$f(\mathcal{R})$ model, but may still be correct for some other
modified gravity theory. Any such theory would, however, likely be
subject to additional constraints from local tests of gravity.

Despite of their similar behaviours, Palatini $f(\mathcal{R})$
gravities and GR are not equivalent. Even though motions of
isolated bodies are the same in both theories, the internal
structure and dynamics of them are generally different
\cite{flan}. This is because, for the body to be stable under
gravity one must require that Eq.~(\ref{volk}) holds, and the
appearance of $\Delta V$ in this equation clearly indicates an
alteration to GR. As we mentioned above, the pressure gradients
inside the particle must be modified with respect to those in GR
in order to exactly cancel the gradients in $\Phi$, and so we
expect the distributions of matter inside this particle, that is,
the internal structures, according to these two theories to be
different.


As another example, the electron energy levels in atoms are
altered in Palatini $f(\mathcal{R})$ theories (see also
\cite{Olmo2008} for a recent work). Such alterations occur,
because in these theories the effective electron mass depends on
the electron density, and so is different for different energy
levels. In the absence of any Palatini modification, we define the
energy and number density of an electron with total angular
momentum $j$, in energy level $n$, to be: $\bar{E}_{nj} =
-\alpha^2 m_{e} \mathcal{E}_{nj}$ and $\bar{n}_{e}^{(nj)}$
respectively where $\alpha$ is the fine structure constant, and
$m_{e}$ the electron mass.  In Palatini $f(\mathcal{R})$ theories,
the electron mass depends on $\Phi$ and hence on the electron
density which is different for different values of $n$ and $j$. We
assume that $f(\Phi) \approx b\Phi (1+\varepsilon(\Phi))$ so that
the density dependence of the electron mass is slight. The total
energy of an electron (up to an overall constant) is then
\cite{longversion}:
$$E_{nj}  = \alpha^2 m^{\rm eff}_{nj}\mathcal{E}_{nj}$$ where
$$m_{nj}^{\rm eff} =
m_{e}(1+\Delta_{nj}/2\alpha^2\mathcal{E}_{nj})$$ for some $m_e$
and to leading order in $\alpha^2$:
\begin{equation}
\Delta_{nj} = -\left\langle \varepsilon(\Phi) \right \rangle_{nj}
\equiv -\int \dd^{3} x\,\bar{n}_{e}^{(nj)}(x)
\varepsilon(\bar{\Phi}_{nj}(\mathbf{x}))
\end{equation}
where $\bar{\Phi}_{nj}(\mathbf{x}) / b = \kappa m_{e}
n_{e}^{(nj)}(\mathbf{x})$. Thus, in Palatini theories
determinations of the electron mass from transitions between
different energy levels using the standard formula for $E_{nl}$
would find different answers for each transition unless
$\Delta_{nl}={\rm const}$; this possibility is however very
strongly constrained. So far both $\varepsilon$ and $m_e$ have only been
defined up to an overall constant \emph{i.e.} $\varepsilon
\rightarrow \varepsilon + \delta_0$, $m_{e} \rightarrow
m_{e}(1+\delta_0)$ for some constant $\delta_0 \ll 1$ is allowed.  We fix the
definition of $m_e$, and hence also $\varepsilon$, so $m_e$ is the
effective electron mass for the ground state \emph{i.e.}
$\left\langle \varepsilon \right\rangle_{10} = 0$. Using measurements of the electron mass from
the transitions 1S-2S and 2S-8D \cite{rydberg} we have
\begin{equation}
\left \vert \left\langle \varepsilon(\Phi) \right \rangle_{20} -
\frac{16}{21}\left\langle\varepsilon(\Phi) \right
\rangle_{83}\right\vert < 8 \times 10^{-16}. \label{bound}
\end{equation}
Respectively for $(10)$, $(20)$ and $(83)$ states the values of
$\rho_{e}= m_{e}n_{e}$ near the expected electron radius are:
$3\times 10^{-5}\,{\rm g\,cm}^{-3}$, $10^{-6}\,{\rm g\,cm}^{-3}$
and $10^{-8}\,{\rm g\,cm}^{-3}$. If, for instance,
$\varepsilon(\Phi) \approx {\rm const} + \epsilon_{0}\Phi/ b
H_0^2$, where $H_{0}^2$ is the value of the cosmological constant
today, Eq. (\ref{bound}) gives the very strong constraint:
$$\vert \epsilon_{0} \vert \approx \vert f^{\prime
\prime}(\Phi)H_{0}^2/f^{\prime}(\Phi) \vert \lesssim 4 \times 10^{-40}$$

In summary, much of our intuition about how the microscopic
behaviour of gravity affects physics on large scales is based upon
Einstein's general relativity. In this \emph{Letter} we show that
such an intuition \emph{cannot} simply be generalized to modified
gravity theories without a detailed analysis of the
energy-momentum microstructure. Indeed, na\"{i}vely averaging over
the microscopic structure will generally lead to incorrect
predictions, and inaccurate conclusions as to the validity of the
theory. In particular, the na\"{i}ve averaging procedure is
\emph{invalid} in Palatini theories. A correct averaging procedure
shows that the cosmology of Palatini $f(R)$ models is identical to
that of GR and fine tuning problems associated with the
cosmological constant are neither alleviated nor, it should be said, worsened.  Furthermore, the relative motion of particles in
Palatini theories is indistinguishable from that predicted by GR.
Interestingly, although Palatini $f(R)$ theories were designed to
modify gravity on large scales, they actually modify physics on
the smallest scales (\emph{e.g.} the energy levels of electrons)
leaving the larger scales practically unaltered. In general,
before considering any astrophysical consequences of a modified
gravity theory, it is important then to check that it does not
make unrealistic predictions for atomic physics.

One may wonder whether similar problems arise in the
metric $f(R)$ gravity theories. In this paper, we have been concerned with
averaging over very small scales 
(e.g. atomic scales). In the metric $f(R)$ theories, the extra scalar degree
of freedom is \emph{dynamical} and so it can and does propagate.  This means that its
dependence on the distribution of matter is not so rigid as in
Palatini theories. Over the very small scales of particles,
although the energy density of matter might change rapidly (from nuclear
density inside the particles to zero outside), the extra scalar
degree of freedom $\Phi$ is not required to change so abruptly. In metric $f(R)$ theories, the field equation for $\Phi$ has the form: $\nabla^2 \Phi
+ V^{\prime}(\Phi) \propto (\rho - 3p)$, where $p$ is the pressure of matter and $\rho$ its energy density.  Over very small scales, the kinetic term ($\nabla^2 \Phi$) dominates over the potential term ($V^{\prime}$) and so reduces to $\nabla^2 \Phi \propto (\rho - 3p)$.    This means that over small scales the leading order deviation from GR, determined by $\Phi$, obeys a linear second order differential equation.   In Palatini theories, $\Phi$ is related
algebraically to $\rho - 3p$. If $\Phi$ depends
\emph{algebraically} and \emph{linearly} on $\rho - 3p$, we reduce
to GR.  It follows that in Palatini theories the leading order
deviation from GR is necessarily determined by a non-linear
algebraic relation to $\rho - 3p$. 

One may think of metric $f(R)$
theories as being scalar-tensor theories with a gravitational
strength coupling to matter, and Palatini theories as being ones
with an essentially infinite strength coupling to matter (in this
sense they also have an infinite mass, but the ratio of the mass
to the coupling is finite).  If one considers a general
scalar-tensor theory with arbitrary strength coupling (but defined
so that the mass divided by the coupling is fixed) then one would
expect to see a cross-over, for some coupling strength, from a
behaviour where the macroscopic dynamics are determined, to
leading order, by the microscopic field equations (and averaging
works as one might expect), to a behaviour where the average
macroscopic dynamics are \emph{not} described by the microscopic field
equations.  This is precisely the behaviour that was found in the
study conducted in \cite{Motashaw}. For a relatively weak
(e.g. gravitational strength) coupling, such as in metric $f(R)$
theories, averaging over small scales works as one would generally expect (at least
to determine the leading order deviation from GR). It should be stressed though that these problems with averaging may re-emerge in metric $f(R)$ on larger scales (e.g. those of large scale structure in the Universe). A discussion of this scenario is beyond the scope of this article. 

Another point worthy of discussion is that the conclusions of this paper concerning the macroscopic behaviour of Palatini $f(R)$ theories applies generally well to much wider class of theories.  Specifically, for a given theory on can always write (for a some choice of conformal frame) the modified Einstein equations as:
\begin{eqnarray}
R^{\mu}{}_{\nu} - \frac{1}{2}R \delta^{\mu}{}_{\nu} + \Lambda \delta^{\mu}{}_{\nu} = \kappa T^{\mu}_{{\rm m}\,{\nu}} + t^{\mu}{}_{\nu}, \label{modEqn} 
\end{eqnarray} 
where $\Lambda$ is some cosmological constant term, $T^{\mu}_{{\rm m}\,{\nu}}$ is the energy momentum tensor of matter and $t^{\mu}{}_{\nu}$  represents all of the modifications from the standard Einstein equation.  All of our conclusions of the macroscopic behaviour of Palatini $f(R)$ theories, will then apply if, for some choice of $\Lambda$, both terms on the right hand side of Eq. (\ref{modEqn}) vanish outside particles. One could then replace $\kappa T^{\mu}_{{\rm m}\,{\nu}} + t^{\mu}{}_{\nu}$ by $\kappa \tilde{T}^{\mu}{}_{\nu}$, and this new energy momentum tensor would then only have support where $T^{\mu}{}_{{\rm m}\,\nu}$ i.e. if $T^{\mu}{}_{{\rm m}\,\nu}$ describes a system of particles separated by vacuum then so does  $\tilde{T}^{\mu}{}_{\nu}$.   In any such theory, one could, as is done in Ref. \cite{longversion}, deduce the
dynamics of particles, and hence the averaged dynamics of a set of
particles, simply by considering surface integrals which depend
only on the form of the metric outside the particles. Outside the particles, the field equations in this modified theory reduce to $G^{\mu}{}_{\nu} + \Lambda \delta^{\mu}{}_{\nu}$ i.e. just vacuum GR. In this modified theory, particles in a vacuum  would therefore move precisely like particles in a vacuum in GR. Since the motion of such particles in the later does not dependent on the precise composition of the particles, it is essentially irrelevant that we have replaced the original matter energy momentum tensor with a modified one.  It is straightforward to see that if the motion of particles is equivalent to that in GR for one choice of conformal frame, it is equivalent for all choices of conformal frame.

\section*{Acknowledgements}
The authors thank G.~Olmo, E.~Flanagan and N.~Kaloper for
discussions. BL, DFM and DJS acknowledge the Overseas Research
Studentship Award, Cambridge Overseas Trust, the Humboldt
Foundation and STFC respectively.

\end{document}